\hsize=155mm   \vsize=215mm
\hoffset=0mm
\baselineskip 20 pt
\overfullrule 0pt
\magnification=\magstep1
\font \mi =cmbx10 scaled\magstep1
\def \dodo  { \partial ^0   \partial _0 }
\def \alp  {\alpha}
\def \sig {\sigma}
\def \nab {\nabla}
\def \gam {\gamma}
\def \noi {\noindent}

\def \hron {{\cal H}}

\def \half  {  {1 \over 2 } } 
\def \eron {{\cal E}}
\def \sron {{\cal S}}
    \def  \defD {2} 
  \def  \iso   {3}    \def  \frw   {4}
 \def  \lapl   {5}    \def  \Reduc   {6}   
  \def  \pectr   {7}       \def  \reduc'   {8}   
  \def  \gensol   {9}     \def  \sesq   {10}   
  \def  \2.1   {11}    \def  \wro   {12}   
  \def  \exep   {13}   \def  \Tpsi   {14}   
  \def  \??   {15}     \def  \LL   {16}   
  \def  \deplus   {17}     \def  \epsi   {18}   
  \def  \kernl   {20}        \def  \gamoto   {21}   
  \def  \invar   {22}      \def  \somu   {24}   
  \def  \biscal   {25}     \def  \sumker   {26}   

\def  \Dew {1}     \def \licihes {2} 
  \def \jap {3}     \def  \moren {4}  
 \def  \birl {5}  
\def \wein       {6}          \def \ouch       {7}             
   \def \pdv       {8} 
\def \seg  {9}          \def \cheval   {10}         \def  \unit {11}  
  \def \sit   {12}          \def \energ    {13}               
 \def  \Sitt   {14}   
    \def  \berg {15}  
   \def \parkful   {16} 
 \def \bel {17}             \def  \ave {18}

\centerline {\mi  \bf  Invariance of Positive-Frequency Kernels}
\centerline  {\mi  \bf in Generalized FRW Spacetimes}
\bigskip
\centerline {\bf Philippe Droz-Vincent}
\bigskip
\centerline    {Laboratoire de Gravitation et Cosmologie Relativistes}
\centerline {C.N.R.S.   URA 769, Universit\'e Pierre et Marie Curie}
\centerline {Tour 22-12,  Boite Courrier 142}
\centerline {4 place Jussieu    75252 Paris Cedex 05, France}
$$ \         $$ 
\bigskip
\noi             
{\sl  We consider the Klein-Gordon equation 
in FRW-like spacetimes, with compact space 
sections (not necessarily isotropic neither homogeneous). The bi-scalar kernel 
allowing to select the positive-frequency part of any solution is developed 
on 
mode solutions, using the eigenfunctions of the three-dimensional Laplacian.
Of course, this kernel is not unique  but,  except (perhaps) when the scale 
factor undergoes a special law of evolution, the metric has no more symmetries 
(connected with the identity) than those inherited from the space sections.
As a result, all admissible definitions of the positive-frequency kernel are 
related one to another by a unitary transformation which commutes with the 
connected isometries of spacetime; any such kernel is invariant under these 
isometries.  A physical interpretation is tentatively suggested\/}.

   03.65.P   $   \qquad     $  04.62

\vfill  \eject

{\bf I. INTRODUCTION.}

\noi  
Having in mind a theory of {\it free} quantum fields in a Lorentzian 
manifold $V_4$, we plan to later undertake the construction of Fock space.
But the first step will be to consider one-particle states;  therefore we write
the wave equation  describing the minimal coupling of a particle with mass $m$
$$ ( \nab ^2  +  m^ 2  ) \Psi =0       \eqno (1)   $$
 for a complex-valued function $\Psi $   of class $C^ \infty $.
The theory of its elementar solutions, satisfying retarded or advanced 
conditions, has been intensively developed [\Dew][\licihes].
For a given manifold they are unique and invariant by the 
isometries of $V_4$ [\licihes].
In the case of Friedman-Robertson-Walker   (FRW) models,
 explicit expressions for these 
propagators  have been calculated [\jap][\birl][\moren].

\noi
Of different nature are the kernels associated with the selection of a 
positive-frequency part in the solutions [\wein].
In spite of being related to the propagator by a convolution equation [\ouch]
they cannot be just derived from the fundamental solutions.
Worse than that,  their definition seems to suffer 
 (except in the stationary case) from  a high degree of arbitrariness.

\noi   In sofar as FRW universes are concerned, it is in principle possible to 
calculate the $G_1$ kernel of Lichnerowicz along the lines of ref [\moren].
But the mathematical framework used in that work, justified if one aims at a 
general theory, is unnecessarily complicated for applications to FRW 
spacetimes, mainly because it disregards the concept of mode solutions.
Therefore obtaining compact and tractable formula will be a new  advance.
In addition we shall be concerned by the question of symmetries.

\bigskip
\noi
We are faced with the problem  of splitting  the  linear space of the solutions 
to equation 
(1) into some generalization of positive-frequency  and negative-frequency
 subspaces [\pdv].
In a popular terminology one speaks of the problem of "defining the  vacuum".
In our opinion it is rather the problem of defining one-particle states.

\noi         The determination of a  positive-frequency
 subspace amounts to the selection of a "complex-structure positive operator"
[\seg][\ouch][\moren],
 that is a 
real linear operator $J$ (acting on solutions) such that $J^2 = -1$.
The action of  $J$ can be expressed with the help of a kernel $G_1$ in terms 
of the sesquilinear form associated with the Gordon current [\ouch].
This kernel can be uniquely selected in the particular case of stationary 
spacetime [\cheval];
 in this case energy (defined as time component of momentum) 
is conserved,  and positive frequency corresponds to  positive energy.
In contradistinction, quantization in {\it nonstationary spacetimes\/}
 is  generally plagued with ambiguities about the determination of 
positive-frequency solutions.

\noi
The lack of unicity in the splitting of the space of solutions is somehow  
redeemed by the well-known fact that two possible candidates as 
positive-frequency subspace can be mapped one onto the other by a unitary 
transformation. Known for  a long time in the FRW case, this property 
has been extended to arbitrary spacetime by the work of Moreno [\unit].
This wide class of equivalence can be further restricted if we introduce 
reasonable symmetry requirements.
We mean for instance that   quantum field theory in any spacetime should 
respect  isometries, as much as it does in the Minkowski case.
In fact a similar concern has been already expressed long time ago
 in the special case of de Sitter space [\sit].

\noi     Notice however that, in special relativity,
 only orthochronous Lorentz transformations 
preserve positivity of energy. Similarly, in curved spacetime, the role of 
discrete isometries should be investigated; but for simplicity of the 
exposition, we shall mainly 
focus on  the isometries that are continuously connected with 
the identity  ({\sl connected  isometry group\/}).
 The possibility of including discrete isometries will be briefly discussed at 
the end  of this work.

\noi
 As most metrics of physical interest enjoy some symmetries, 
we shall complete the axiomatics of ref. [\ouch] by a condition of 
{ \sl isometric invariance}.   

\medskip
\noi
Technically, we prefer to deal with the positive-frequency  
and  negative-frequency  projectors  
$ \displaystyle   \Pi ^\pm =   {1   \pm  i J  \over  2 } $
  rather than with  $J$.  This notation amounts to consider the 
positive-energy and negative-energy kernels
  $D ^ \pm  (x, y) $                            instead of  $G_1$. 
They must be distribution solutions to  the wave equation (1) and
 satisfy the formula
$$  (\Psi ^\pm) (y) =  (D^\pm _y  ; \Psi )      \eqno (\defD )     $$
where  $\Psi ^\pm  =  \Pi ^\pm    \  \Psi $
 is the positive-frequency part of $\Psi$, 
$y$ is an arbitrary point of $V_4$,
and  $ (\Phi ; \Psi ) $     denotes the sesquilinear form constructed with the 
help of the Gordon current, conservative provided $\Phi $ and $\Psi $ are 
solutions to the KG equation.
Our interest for this approach is motivated by  the fact that 
 the field operator must be defined through  the creation and anihilation 
operators   associated  with  the one-particle state  $D^+$.
In terms of this  kernel, we shall formulate isometric invariance as follows
  $$ D^+  (x, y)   =  D ^ + ( Tx , Ty)     \eqno (\iso )         $$
for any metric-preserving transformation $T$ of the connected isometry group.
When this property is satisfied one can reasonably expect that field 
operators, commutation relations, and the whole machinery of Fock space 
construction will be isometrically  invariant.                Of course,
 the condition of isometric invariance substancially reduces the arbitrariness 
in the choice of $J$, without fixing it completely.

\medskip
\noi    Our purpose is now to prove that, for a large class of nonstationary 
spacetimes,  all resonable splittings of the space of solutions are mutually
 equivalent, {\it in a way which respects the symmetries of free motion\/}, 
and more specially isometric invariance.

\medskip
\noi   In this paper we shall consider a simple form of nonstatic orthogonal
 spacetime, large enough to encompass ordinary FRW universes
 (without singularity) as a particular case. 
In analogy with usual FRW universes, the spacetimes we deal with 
admit, in the generic case,  distinguished  timelike curves and space sections.
In order to obtain rigorous results we shall eventually 
assume that $V_4$ can be identified 
with ${\bf R } \times V_3 $ where $V_3$ is a connected, compact,
 three-dimensional space.
In contrast to the usual FRW models, neither  isotropy nor homogeneity of space
 sections is necessary here.

\noi   All the space of the form we consider have interesting features:

\noi   $i)$  The wave equation can be reduced to an ordinary (second order) 
 differential equation, complemented by the eigenvalue equation for the 
three-dimensional Laplacian. 
This reduction stems from the conservation of a (purely 
spatial) generalization of kinetic energy.

\noi $ii)$  
At least in the generic case,  the isometry group is under control.

\medskip
\noi
For  spacetimes  of this kind , the dynamical symmetry group countains not 
only isometries but also the 
transformation generated by "kinetic energy".

\noi
The main achievement of this article states that
 the demand of isometric invariance, associated with the natural requirement
 that the separation of frequencies respects kinetic-energy shells,     
     permits   us to select 
 a unique kernel $D^+$,  {\it  up to a unitary transformation 
 commuting with all connected isometries\/}.
This result is in the line of the main theorem proved in ref.[\moren]; 
but the work presented here goes one step further
 by taking the role of symmetries into account.
In [\moren] explicit formulas concerning the Lichnerowicz kernel were 
displayed.  
Here, the expansion  we  obtain directly concerns  the positive-frequency 
kernels  $ D^ + $.
Strictly speaking it is not a closed form expression, but it is resonably 
compact and its  invariance  properties can be easily read off.

\noi    Notice also that   Moreno  in ref.[\moren] treated the Klein-Gordon 
 equation  as  an evolution equation, 
and this was probably a technical necessity for his  purpose.
Here  we avoid this complication, unnecessary in the simple case of
FRW-like spacetimes. 

\medskip
\noi   in Section II, the intuitive feeling that "in general" FWR spacetimes 
have no more symmetries than those provided by the group of motion in space 
sections is confirmed and given a precise meaning.

\noi  We write down in Section III,   for the $D^+ $ kernels, a
 tractable  expansion over modes and over the eigenfunctions of the 
three-dimensional Laplacian. We  explicitly check that this
 expression  is invariant under all the isometries of  the space sections. 

\noi  Finally the unitary equivalence of all  admissible 
definitions of $D^+$ will be exhibited, in a way which respects  
 all the  isometries of $V_3$,  ensuring by the same token (in the generic 
case) invariance under all the transformations of the {\it connected} 
isometry group of  of $V_4$.

\noi     Section IV is devoted to a few comments and  a tentative 
interpretation.

\bigskip
\noi
 {\bf II.  GENERALIZED  FRW-SPACETIMES}

\noi     {\bf II.1 Generalized kinetic energy}

\noi
We assume  that the spacetime manifold is 
$ V_4  = {\bf R } \times  V_3 $ and for some $t$
$$ ds^ 2 = 
    B^ 6 dt^2 - B^ 2 d \sig ^ 2 
\eqno (\frw)   $$
where  $B$ is a strictly positive function of $x^0 =t $ and 
$ d \sig ^2 =  \gam _{ij}( x^ k ) dx^i dx^j  $
 defines an elliptic metric.
Notice that
  $ (V_3, d \sig^2 ) $
 has arbitrary curvature.

\noi
We could equivalently write 
$ ds^ 2 =  d{\overline t}^ 2 - B^2 d \sig ^ 2  $, 
if we were to use  the cosmic  time $\overline t$.

\noi
However, unless otherwize specified,  we shall stick to the form (\frw) of
 the metric because it eliminates first derivatives from the reduced wave 
equation and leads to a simple formula for the Gordon current, implying 
 equation (\wro ) below.

\noi
In spacetimes of the above type, the dynamics of a free particle enjoys an 
interesting property.
Indeed {\sl free motion in these spacetimes admits a first integral
 which generalizes kinetic energy} [\energ].

\noi
This can be  checked already at the {\it classical}
 level using a covariant symplectic Hamiltonian framework.
 In this formalism, phase space is the cotangent bundle  $T_*(V_4) $
endowed with coordinates $x^\alp , p_\beta$ submitted to the standard Poisson 
bracket relations.

\noi     Geodesic motion is generated by the Hamiltonian function
$$ H= {1 \over 2} g ^{\alp \beta} p_ \alp  p _ \beta    $$
Taking  (\frw) into account and defining
$$ 2K =   \gam ^{ij} p_i p_j   $$
where  $\gam$ with latin superscripts refers to the inverse tensor of
 $\gam _{ij}$, say         
$    \gam ^{ij}   \gam _{jk} = \delta ^i _k   $,
  we easily obtain 
$$ 2H = B ^{-6} p_0  p_0  - 2 B^{-2} K      $$
Since  $ \{x ^i , p_0 \} = \{x^0 , p_j \} = 0 $  it follows that $K$ has a 
vanishing Poisson bracket with $ B^{-6} p_0 p_0 $ and $B^{-2} $ hence 
finally also with $H$, which proves our statement.
Notice that $K$ involves space variables $x^k , p _j $ only.

\medskip
\noi

\noi     The {\it quantum\/}
 mechanical version  of  $K$ is proportional to the 
three-dimensional Laplacian. Its commutation  with the Klein-Gordon operator 
permits to separate variables in the wave equation (see (\gensol ) below).

\noi  Notice that the quantum analogous of $H$ is nothing but 
$  \displaystyle  -{1 \over 2} \nabla ^2 $.

\noi
When $V_4$ has no more isometries than those corresponding to symmetries of 
$V_3$,   it is clear that  
$K$ is invariant under all isometries of $V_4$ (unless otherwize specified we 
consider the connected group).

\noi
Beside these obvious symmetries,  a  question is wether additional  isometries
 may exist in $V_4$.

\noi      This may happen.
For instance let us  consider  a FRW universe in 
the usual sense, with maximally symmetric  space sections.
Any transformation taken from the group of motions in the 
 three-dimensional space corresponds to  an isometry of spacetime.
We expect that "in the generic case", the metric has no further symmetry.

\noi                           But  if  the  universe expands in a 
particular way, it may be  a de Sitter manifold; then the group of  
isometries has   ten parameters. The group of  motions in $V_3$
    do not exhaust all
 the  isometries of $V_4$, but it might be reasonable   to require 
that the Weinberg function be  invariant under all of them. 
Since quantization in de Sitter space has been extensively discussed in the 
literature [\sit][\Sitt],  we shall not dwell further in this direction.

\noi
Owing to the importance of isometries in the problem of quantization, it is of 
interest to make sure that the  situation pictured in  the above example 
is in fact exceptional.
This point will be clarified in  subsection II.3.

\medskip
\noi
 {\bf II.2  Separation of variables in generalized FRW spacetimes.}

\noi For the moment eq.(\frw) allows one  to cast (1) 
into a remarkably simple form. We use coordinates such that 
$$    g^ {0i} = g_ {i0}  =  0  \qquad 
   g_ {ij} =  -B^{2} \gam_ {ij}          $$ 
$$ \det g_{ij} = B^ {6} \gam  ,\qquad   \gam  = \det \gam_ {ij}  $$
 $ g = \det  g_ {\mu \nu} ,  \quad$  
         $ \sqrt {|g|} = B^ {6} \sqrt \gam  $.

\noi
It turns out that  $\nab^ 2$ and therefore the operator in (1)
 commute with $\Delta _3$. This fact  permits 
 to reduce the wave equation into a one dimensional problem. 
Solutions to  the wave equation exist which are also eigenstates of 
the three-dimensional Laplacian $\Delta_3$.
In accordance with the usual terminology applied to Friedman universes, we 
shall call   them  {\sl mode solutions} of eq. (1).

\noi       Let us  consider a mode solution, say $\Phi$. 
For some nonnegative $\lambda \in Spec(V_3) $ we have
$$ \Delta_3  \Phi =  - \lambda \Phi     \eqno (\lapl )  $$
and (1) reduces to
$$ (\dodo +  \lambda  B^ {-2}  + \mu   )  \Phi  = 0   \eqno (\Reduc)  $$ 
with   $  \partial ^0  = B^{-6} \partial _0       $.
The space variables $ x ^ j$ can be ignored in solving this equation.
Since  the coefficient of
 $ \partial _  0  \partial _0  $ 
in $\dodo$ never vanishes,   equation  (\Reduc) is always of second order.
Here we assume that $V_3$ is compact and simply connected, which entails 
[\berg][\parkful]  that     $\lambda$   belongs to the discrete   sequence 
$$  Spec(V_3)  = 
    \{        \lambda_0 = 0 \leq  \    
\lambda_1 \leq  \  \lambda _2 \leq  \   \lambda_3 \  .....   \infty   \}
                                                      \eqno (\pectr)  $$
and that  the eigenspace  
${\cal E}_n$  of  $- \Delta _3$  in $  C^\infty (V_3) $,  associated 
with the eigenvalue  $\lambda _n $  has finite dimension      $r (n)$.

\medskip
\noi                      Let $ {\cal S}_ n  $ 
be the space (two-dimensional over complex numbers)
of $C^ \infty $ functions of $t$ satisfying the equation
$$  ( \dodo +  \lambda _n   B^ {-2}  + \mu   )  f = 0    \eqno (\reduc' )   $$
and let 
$ f_{1 ,  n } (t) $ and 
$f _{2 , n  } (t) $      
form a basis of    $ {\cal S}_ n   $.
The general solution of (1) in mode $n$ can be written as 
$$ \Phi =   F_{1,n} ( x^ j)  f_{1,  n } (t)    +
  F _{2,n}  (  x^ j)     f_{2,  n } (t)       \eqno (\gensol )   $$
where $F _{1,n}$ and  $ F_{2,n}   $ belong to   $\eron _n$.

\noi     {\sl Definition: 
 Any basis of }       $ {\cal S }_ n   $ {\sl such that} 
$ f _{2, n}  =  f _{1, n} ^ *  $ {\sl will be called a  canonical
basis}. 

\noi
Remark: starting from two independent real solutions of (8) it is 
always possible to construct  a canonical basis.
  Each such basis permits to define in   $  {\cal S} _ n$  the one-dimensional 
subspaces   $   { \cal S}_n ^ {(1)} $  and 
$   { \cal S}_n  ^ {(2)}    $ 
spanned by  $f_{1, n}$ and $f_{2, n} $ respectively.

\noi     {\sl Definition:
A canonical basis $f_{1,n} , \   f_{2,n} $ is orthonormal when 
$ i W (f_1 , f_2 ) = 1 $\/}.

\noi   Here the Wronskian 
 $W(f, h) \equiv  f \dot h - h \dot f$             is constant in $t$ 
provided both $f$ and $h$ are solutions of the same equation (\reduc' ).

\noi
Our terminology is justified by the fact  that $-i W(f^* , h) $ is a 
sesquilinear form of the couple $f,h$ in $\sron _n$.

\medskip
\noi    It is natural to postulate that the projectors $\Pi ^\pm $ commute
 with $K$.
This property is  always implicitly assumed in the literature.
It amounts to require that the splitting into positive and negative-frequency 
subspaces is first performed in each "kinetic-energy shell" (eigenspace 
common to $\nabla ^2 $ and $ \Delta _3$) and further extended by direct sum.
As a result,  the only freedom left for defining positive-energy  solutions 
concerns the splitting of 
$\sron _n S $   into    $ {\cal S}_n  ^{(1)}$  and    $ {\cal S}_n  ^{(2)}  $.
One must proceed so in  each kinetic-energy shell, 
but in this section the eigenvalue $\lambda _n$ is kept fixed.
Whenever no confusion is possible, we shall drop the label $n$ referring to it.

\noi  In order to discuss the  surviving degree of arbitrariness,
 let us consider two admissible choices respectively characterized by 
  $f_1, f_2$  and another canonical basis, say
$ f '_1 ,  f ' _ 2 $.

\bigskip
\noi   
{\sl The  sesquilinear  form}
$$ (\Phi ; \Omega) =  
\int j^\nu (\Phi, \Omega) d\Sigma _\nu                 \eqno  (\sesq )       $$
associated with the Gordon current 
$$ j^\nu (\Phi , \Omega) 
 = -i ( \Phi ^*  \nab ^ \nu \Omega -  \Omega \nab ^\nu \Phi^ * )
                                                    \eqno (\2.1 ) $$
is independent of the spacelike hypersurface $\Sigma$
   for any couple of solutions to (1).
For product solutions   of the form
 $  \quad  \Phi = f(t)  F( {\xi } ),   \qquad  
\Omega =  h(t)  H({\xi  })  \    $,
namely for mode solutions,
our choice of time coordinate permits to write this simple formula 
 $$  (\Phi; \Omega) =  
\int_ \Sigma j^ \nu (\Phi, \Omega) d \Sigma _ \nu =
-i  W (f^ *, h)  \int F^ * H \sqrt \gam  d^ 3 {\xi  }  \eqno (\wro )    $$
The integral in the r.h.s. is the three-dimensional scalar product  
$$  ((F, H )) =      \int F^ * H \sqrt \gam  d^ 3 {\xi  }    $$
invariant by the isometries of $V_3$ endowed with metric $\gam$ 
({\it spatial } isometries).

\noi
Notice that formula (\wro ) ensures {\it positivity\/} of 
$    (\Phi ; \Omega) $ provided  the basis of 
${\cal S}$        is suitably normalized.
We can always choose our notation such that 
$ -i W(f_2, f_1) = 1 $.
 As seen in a previous work (Propo.3 in ref.[\pdv])
we can assert that $ (\Phi  ;  \Phi)  \geq 0 $ and  vanishes only for 
$\Phi =0$,       when
 $\Phi  \in {\cal S}^{(1)}   \otimes {\cal E} $. 
Similarly  $ (\Phi  ;  \Phi)  \leq 0 $ and  vanishes only for 
$\Phi =0$,       when
 $\Phi  \in {\cal S}^{(2)}   \otimes {\cal E} $. 

\noi
The {\sl scalar product\/} is defined as  $ <\Phi , \Psi >= \pm (\Phi ; \Psi) $ 
respectively in  $ \sron  ^{(1)} \otimes  \eron $  and 
        $ \sron  ^{(2)}  \otimes \eron $,
whereas      $ \sron  ^{(1)}  \otimes \eron $               and 
        $ \sron  ^{(2)} \otimes  \eron  $                    are 
mutually orthogonal in the scalar product  as well as they are in the 
sesquilinear   form (\sesq ).
In contradistinction to this sesquilinear form, the scalar product defined  
above is positive definite, hence Hilbertian, for $\dim \eron $ is finite.
But it crucially depends on  the splitting we have performed,
 that is on the choice of an orthonormal canonical  basis in $\sron$.

\bigskip
\noi    {\bf II.3 Isometries.}

\noi  Generalized FRW spacetimes have obvious symmetries:
It is clear that any isometriy of $V_3$ also leave invariant the spacetime 
metric.
 But the question is to investigate if  other isometries can occur in $V_4$.

\noi  {\bf \sl Proposition 1.}

\noi
{\sl Except perhaps in the case where the scale factor satisfies a particular 
law of evolution,  
all connected isometries of $V_4$ are induced by those of  $V_3$}.

\noi  The proof is purely local; it essentially involves infinitesimal 
generators.
  Let $X$ be a Killing vector on $V_ 4$. When it corresponds to a group of 
motion in $V_3$, its component $X^0$ vanishes.
Otherwize, we shall say that $X$ generates a "nontrivial group of isometries".
 But is this latter 
case really possible?  The answer is {\it no\/} in the generic case precisely 
characterized below.

\noi
Calculations are carried out in Appendix, using the  cosmic  time 
coordinate, and setting $S=B^2$. 
It is clear that the vanishing of $X^0$ is not affected by any 
 rescaling of the form        $t = t(\overline t)$.  

\noi Starting from the Killing equation one finds  that $X^0$ 
 necessarily vanishes, except if  $S$ satisfies
$$      {d^2 S     \over  d \overline t  ^2} - 
 S^{-1}  ({d  S  \over  d{ \overline t}}  )^2      = const   \eqno (\exep )  $$
In the sequel,  we make the convention that a {\it  generic\/} scale factor 
{\it does not\/}  satisfy the above equation.

\bigskip

\noi    {\bf \sl  Proposition 2.}

\noi {\sl In the generic case, all the connected isometries of $V_4$ are 
naturally represented  as unitary transformations of 
 $\sron  ^{(1)}  \otimes \eron $    
  endowed with  the scalar product $<\Phi , \Psi>$}.

\noi   Indeed we know from Proposition 1  that they are provided  by spatial 
symmetries.  Let $T$ be such a symmetry. It maps  $\eron $  onto itself.
We shall first define  $T F = F (T \xi)$.
Then $T$ is extended to           $\sron  ^{(1)}  \otimes \eron $ as follows.
Say  $\Psi = f _1  (t)  \  F(\xi)$. We write
$$T \Psi =  f_1 (t) \   F( T \xi)                  \eqno  (\Tpsi )        $$
Obviously $T$ maps    $\sron  ^{(1)}  \otimes \eron $  into itself.
It is clear that $T$ is unitary in $\eron $ endowed with the elliptic scalar 
product  $(( F,G))$, obviously invariant under the isometries of $V_3$.
Then we find    $  <T \Phi ,  T \Psi >    =  <\Phi , \Psi>  $
with help of formula (\wro).  Since $T$ is supposed to belong to a continuous 
group it is invertible, which achieve to prove unitarity.

\noi  Of course we have the same property with respect to
  $\sron ^{(2)}    \otimes  \eron$. 

\noi From now on we respectively identify 
 $$    \hron _n ^{(1)} =
\sron ^{(1)} _n   \otimes  \eron  _n ,  \qquad
  \hron _n ^{(2)} =
\sron ^{(2)} _n   \otimes  \eron  _n                                $$
as the positive- and negative-frequency subspace in the mode labelled by 
$n$.          According to  (\gensol), the $n^{th}$ space of mode-solutions is
$$ \hron _n = \hron _n ^{(1)}   \oplus    \hron _n ^{(2)}              $$
In the sequel we shall indifferently write  
$\hron _n ^+ $ for   $\hron _n  ^{(1)}$ , etc.

\bigskip
\noi              {\bf III.  POSITIVE-FREQUENCY KERNEL}

\noi  {\bf III.1 Kernel at a given mode}.

\noi
We turn back to quantum mechanics and consider equation (1) when the metric 
is of the form (\frw).
Separation of frequencies is supposed to respect kinetic-energy shells, 
therefore in each shell $ \hron _ n $ we must have  the projectors
  $\Pi ^\pm _n $ associated with the kernels  $D^\pm _n$.

\noi
But kinetic energy is kept fixed throughout {\it this section\/}; therefore,
 {\it the label $n$ referring to a determined eigenvalue
 $\lambda _n$ is dropped in  intermediate calculations\/}.

\noi  
In this section we assume that some canonical orthonormal 
basis of ${\cal S} $ is choosen.        Let it be  $  f _1  ,  f_2 $ 
with    $f _2 = f _1 ^* $. Being orthonormal it satisfies 
 $ -iW (f _2,  f _1  ) = 1  $.

\noi   It is required  that  $D^ + $  itself is
  (possibly in the sense of distributions)   a positive-frequency solution 
of (1).
Therefore the kernel $D^+ _n (y, x)$, as a function of $x$, must belong to
   $\sron ^{(1)} _n   \otimes  \eron  _n $. In other words, with an obvious 
notation 
 $  y= (u, \eta) , \quad  x = (t, \xi) \  $, 
we can write 
$$ D^+  _n    (y,x)  =  f _1 (t) \   L (y, \xi)   \eqno (\?? )    $$
where  $L $, as a function of  $\xi$  
 must belong to $\cal E$. 
 Naturally, owing to the bi-scalar nature of the 
kernels,    $L$ may additionally  depend on $y$.

\noi
Here we take advantage of the fact that  $\dim \  {\cal E} =r(n)<  \infty $.
Let $  E_1, E_2, .... E_r $ be an orthonormal  basis of $\cal E$, that is 
$ ((E_a ,E_b )) = \delta _{ab}$.
We can write  $L $ as 
$$L =  w_1 E_1  +  ...w_r  E _r           \eqno (\LL ) $$ 
with $y$-depending coefficients.
Hence    
   $$ D^+(y,x)  =  f _1 (t) \  \sum w_b E_b (\xi )     \eqno (\deplus)       $$
Consider any  positive-energy solution  $\Psi$ in mode $n$.
It has a development 
$$ \Psi = \sum  f_1 (t) \psi _a E_a (\xi )      \eqno (\epsi )      $$
where $ \psi _1 ,   ....\psi_r $ are complex constants.
According to (\defD ) we must have 
$$ ( (D^+_n)_y ;  \Psi ) =   \Psi (y)          \eqno (19)              $$
where the notation  $  (D^+_n)_y $ indicates that the sesquilinear form is 
calculated by integrating over $x$, and that $D^+ _n$ additionally depend on 
$y$.
According to the developments (\deplus )(\epsi )  we find 
$$ ( D^+ _y ; \Psi )   =          \sum w^* _a  \psi _b    \
(f_1 (t) E_a (\xi )   ;  f_1 (t)  E_b (\xi )  )     $$
Use  (\wro ) and remember that $-i W (f_1 , f_1 ) = 1 $, we get
 $$ ( D^+ _y ; \Psi ) = \sum  w^* _a  \psi _b  (( E_a ; E_b ))     $$
 $$ ( D^+ _y ; \Psi ) = \sum  w^* _a   \psi _a    $$
This quantity must coincide with $\Psi (y)$,  given by  (\epsi ) where 
$t, \xi$   are replaced by  $u, \eta $.
In all this the coefficients $\psi _a$  are arbitrary.
Therefore it is necessary that 
$$  w_a ^* (y) = f_1 (u)  E_a (\eta)               $$
and the only possibility is 
$$ D ^+ (y,x)  =  f_1 ^* (u)  f _1 (t)
           \      \sum E_a^* (\eta) E _a (\xi)       \eqno (\kernl )       $$
In other words,   $D^+ =  f_2 (u)  f _1 (t)  \   \Gamma (\eta, \xi)  $, 
if we set 
$$  \Gamma (\eta, \xi)  =   \sum E_a^* (\eta) E _a (\xi)    \eqno (\gamoto )  $$
It is not difficult to check that  expression (\kernl) of $D^+_n$
actually  implies eq.(\defD ) as it should. 
Moreover  the formula (\gamoto ) 
is invariant  under  changes of   orthonormal basis inside $\cal E$.  
Therefore $\Gamma$ is intrinsically defined; it  is a purely 
spatial quantity, independent of the splitting performed in $\sron$.

\noi  Notice that $\Gamma$ is a reproducing kernel in $V_3$, for we check
$   (( \Gamma _\eta ,  F ))= F(\eta) $ 
for all                           $F (\xi) \in   C^\infty (V_3) $. 
Moreover, using a basis of {\it real functions}  in $\eron$, we see that 
$\Gamma ^* = \Gamma $.

\noi               In contradistinction  the factor   
$   f_1 ^* (u) f_1 (t) $
 depends on the choice of a basis in $\cal S  $. 
For usual FRW spacetimes
 ($V_3$ is of constant curvature), the explicit form of $E_1, ... E_r $ can be 
found in the literature; see references [\berg][\parkful].

\noi As expected we check that  
$ \displaystyle       {D^+ _n (y,x)}^*  =  D_n ^+ (x,y)  $.

\medskip
\noi     Looking for isometric invariance  we can claim

\noi     { \sl    $\Gamma $
is  invariant under all the isometries of $V_3$\/}.

\noi
Proof:

\noi     Let  $T$ be an isometry of the metric  $\gam _{ij}$, we also denote
$T F = F(T \xi)$.
Since transformation $T$ leaves
 invariant the three-dimensional Laplacian,  the  eigenspaces  $\cal E $ are 
globally invariant.
Moreover $T$ leaves invariant the  three-dimensional scalar product,
in other words
$$  (( TF , TG )) = ((F,G))  $$
for  $F, G  \in   C^\infty  (V_3) $.
This is true in particular when $F$ and $ G \    \in \cal E$.
Since $T$ is taken from a continuous group it is inversible;
it follows that $T$ is a unitary transformation of $\cal E$ endowed with its 
scalar product above.
Thus  $ E_1 (T \xi) , .... E_r (T \xi ) $ 
is another basis of $\cal E$. We can write 
$$ E_a (T \xi) =  \sum  T_{ab} E_b (\xi)  $$
hence $$ E^* (T \eta) =  \sum   T^*_{ac} E^* _c (\eta)  $$
Now consider 
$$ \Gamma (T \eta , T \xi ) =  \sum  E^* _ a (T \eta) E _a (T \xi)   $$
$$ \Gamma (T \eta , T \xi ) = 
                 \sum  T^*_{ac} E^* _c (\eta)  T_{ab} E_b (\xi)     $$
But   $ T^* _{ac} = (T^{-1}) _ {ca} $ because of unitarity.    Thus
$ \Gamma (T \eta , T \xi ) =   \Gamma (\eta , \xi)    $.

\medskip
\noi
Now return to (\deplus ). 
 Transformations in $V_3$ do not affect $ \cal S $ and 
leave $f_1, f_2 $  unchanged.
 Proposition 1   ensures that any connected isometry of $V_4$ stems from  an 
isometry of $V_3$ and therefore  $D^+$ is invariant as well as $\Gamma$; 
restoring the mode label we shall write
$$ D^+_n   (Ty , Tx) =  D^+ _n  (y,x)      \eqno  (\invar )  $$
We shall summarize:

\noi                          {\sl The only kernel $D^+ _n $ 
solution of the wave equation, eigenfunction of $-\Delta$ for eigenvalue 
$\lambda _n$
 and satisfying   (\defD ) where $\Psi$ is a mode solution,   is given by 
(\deplus ). In the generic case, it is isometrically invariant.}

\noi  Of course, in the exceptional case it remains at least invariant under 
the subgroup of purely {\it spatial\/} isometries.

\medskip
\noi
Now the next point consists in showing that two possible candidates
 for $D^+ _n$    are necessarily connected by  a unitary transformation
 which respects isometries.

\bigskip
\noi           By (\kernl )(\gamoto )
 it is clear that any alternative candidate for  $D^+$ is 
necessarily of the form
     $$ {D'}^+ =  {f'} ^* _1 (u) \  f' _1 (t) \  \Gamma (\eta , \xi)     $$
where  $f' _1 , f' _2 = {f'}^* _1 $
 form another normed canonical basis of $\cal S$.
We can write 
$$ f'_1 = \alpha f_ 1 + \beta  f_2  $$ 
with $\alpha,  \beta  \in  {\bf C} $. Orthonormality of the new basis reads  
$ \alpha \alpha ^* -  \beta \beta ^* =1$ as is well known.
This  change of basis can be seen also as a map of  $\sron$ onto itself 
(Bogoliubov transformations).
Restricted as a mapping  of 
  $ {\cal S} ^{(1)} $ onto  $ {{\cal S}'}^{(1)} $, it is unitary, say
$f'_1 = U f_1$.
Then $U$ is extended as a unitary map from  
$  {\cal S} ^{(1)}  \otimes  \cal E$  to 
 $ {{\cal S}'}^{(1)} \otimes  \cal E$   ,            according to the rule
$ U (f F) = (Uf) F$.
Now, noticing that  $D^+ _y    \in  {\cal S}^{(1)} \otimes  \cal E $  as a 
function of $x$, we can write {\it in each mode\/}
  $$ ( D'^+ _n)_y   =  U_n  (D^+_n) _y            \eqno  (23)      $$

 \bigskip
\noi   
It is a trivial point that spatial isometries, understood as transformations 
acting in  $\eron _n \otimes  \eron _n $, commute with $U_n$.

\medskip
\noi  Up to now we have proceeded with kinetic energy fixed.
 The last step consists in summing over all the possible values of the integer 
$n$ labelling the eigenvalue $\lambda$.

\bigskip

\noi      {\bf III.2  Sum over modes}

\noi   For each mode, 
$  { \cal H}  _{n} ^ {(1)}$   is equipped with the metric $  <\Phi, \Phi>$
  and   $     { \cal H}  _{n} ^ {(2)}$
with the opposite.
Being of finite dimension  $ {\cal H} ^{(1)} _{n}$  and
    $ {\cal H} ^{(2)}_{n} $ are complete  as Hilbert spaces. 

\noi                        The direct sums 
$   {\cal K}  ^ {(1)}=  \bigoplus   { \cal H} _{n} ^ {(1)}$ and 
$   { \cal K} ^ {(2)}  =  \bigoplus  { \cal H} _{n} ^ {(2)}$ undergo  a 
decomposition wich generalizes the separation of frequencies available in 
static spacetimes. Obviously    
  $ \displaystyle         { \cal K} ^ {(1)}  \oplus
  { \cal K} ^ {(2)} =  \bigoplus    { \cal H} _{n} $.
\noi                    The infinite sum
$    \displaystyle   \Phi = \sum _0 ^\infty     \Phi _n $
where  $\Phi _n  \in   \hron _n $, is in  ${ \cal K}$  only when
$\sum  <\Phi _n  ,  \Phi _n>^2    $ is finite,  but $\Phi$ 
always  exists in the distributional sense, 
if we define as {\sl test functions\/} 
the  sums   $\Psi = \sum \Psi _n $ having   an   arbitrary but 
{\it finite\/} number of terms in  $\hron _n$ (terminating sums).
Of course  $D^+ =  \sum  D^+_n$ exists as a distribution  in the above 
sense (in both arguments).

\noi   When $U_1, U_2 ....U_n,....$  is a sequence such that every $U_n$ acts 
unitarily in $\hron_n$, define 
$$ ( \sum U_n )(\sum \Phi _n) =  \sum  U_n  \   \Phi _n    \eqno (\somu ) $$
It is clear that    $U = \sum U_n$ is a unitary operator in ${\cal K}$.
Moreover $U$ maps into itself the space of terminating sequences 
therefore {\it it can be extended to distributions\/}.

\noi  The above definition applies to bi-scalars with respect to both arguments,
say
$$  (U)_x   (U)_y   \   G = \sum (U_n)_x (U_n)_y  G_n (x,y)  
 \eqno (\biscal ) $$
where  $ G = \sum        G_n$ may  be  a distribution  whereas   
 $G_n  \in   (\hron)_x \otimes  (\hron) _y   $.

\noi
Let  us now take for each $U_n$  the transformation $T_n$ induced in $\hron_n$
by a connected isometry of $V_4$.
\noi  Isometric transformations are extended to the full space of solutions by
equation (\somu).  Again this $U$ maps into itself the space of terminating 
sums, so it can be applied  to distributions.

\noi  It follows from (\invar) and the above formula that 
$D^+ $ is invariant under $(T)_x (T)_y $.

\noi 
The most general  transformation relating two different definitions of 
${\cal K}^{(1)}$ is given by 
$U = \sum  U_n $ where     now each $U_n$
is an arbitrary  Bogoliubov transformation at mode $n$.

\noi Remark: a high energy cut-off (imposing  that  all $ \Phi _n $
vanish for $n$ greater than some fixed integer) would respect isometric 
invariance.
                                        
\noi  We summarize this section in the following statement:

\noi   The positive-frequency kernel  is given by
$$  D^+ (x,y) = \sum _{n=0} ^\infty   D^+ _n (x,y)
\eqno (\sumker)          $$
where  $D^+ _n $ takes on the form (\kernl) in each kinetic-energy shell.

\noi     {\bf \sl Proposition 3}

\noi   {\sl The only kernel solution of the wave equation, mode-wise defined 
 and satisfying (\defD) is given by  the above formula. 
It is defined up to a unitary  transformation
 $U = \bigoplus U_n$  where each $U_n$ is an arbitrary Bogolubov 
transformation in mode $n$.

\noi          In the generic case, $D^+$ is invariant  under the connected 
isometries of $V_4$.}
                          
\noi "Mode-wize defined"   means that equation (\defD) is satisfied mode by 
mode,  and refers to the existance of $D^+_n$ for all $n$.

\bigskip
\noi         {\bf IV.  CONCLUSION}

\noi    The splitting of the space of solutions into positive-frequency and 
negative-frequency subspaces has been carried out separately in each 
kinetic-energy shell, and further extended to the whole space of solutions.
As expected from previous works  on the subject,
 the only ambiguity (for a given value of kinetic energy)
 arises in the reduced space of solutions, 
which is two-dimensional, and corresponds to the possibility of  performing an 
arbitrary Bogoliubov transformation.
Indeed the expansion (\kernl) of $D^+$ in terms of the eigenfunctions of 
$\Delta _3$  still depends, for each mode,
 on the choice of a solution to the reduced (one-dimensional) wave equation,
 restricted only by the Wronskian condition. 

\noi As a result,  there are infinitely many possible  definitions of the 
one-particle space.

\noi   However the total  kernel $D^+$ obtained by summing over modes  
remains  manifestly invariant under the isometries of  space sections. 
 Having checked  that,  beside an   exceptional case,
$V_4$  has generally no more connected isometries  than those inherited 
from its space sections,
  we are in a position to claim that the connected group of spacetime 
isometries  leaves the positive-frequency kernel invariant, irrespective of 
the choice made among all possible candidates.
\noi    Notice that  the (unitary) transformation connecting  two possible 
 definitions of   $D^+$   commutes with connected isometries.

\noi    When $V_3$ enjoys discrete symmetries, which is the case in 
conventional FRW universes ($V_3$ being the three-dimensional sphere),
it is  a mere exercise to check that these symmetries leave $D^+  _n$
 invariant and that they act  unitarily in each $\hron _n ^+$.
Extension to $D^+ $ and ${\cal K}^+$ is straightforward.
But of course these discrete {\it spatial} symmetries may fail to exhaust the 
group of discrete isometries of $V_4$.
There is no statement analogous to Proposition 1  for discrete 
transformations. In fact there are models where $V_4$ is time-reversal 
invariant  whereas, like in Minkowski space, time reflexion transforms 
 $D^+$ into a  negative-frequency solution of the wave equation.

\medskip 
\noi
The technical results presented here are formulated in a self-contained 
manner within the single-particle sector. It is noteworthy that the wronskian 
condition arises as a normalization condition without need to invoke, at this 
stage, the  canonical Hamiltonian formalism for fields.

\noi       Next step will be a Fock space construction and 
the definition of  field operators.
Such a definition  essentially involves nothing
 but the creation and anihilation operators associated with $D^+$.   We 
can anticipate that the invariance of $D^+ $
 will be naturally extended to the whole theory of free fields;
 this point is important because isometric 
invariance plays in  curved spacetime the same role as Poincar\'e invariance 
in special Relativity.

\medskip
\noi  We have left aside the special  cases  where extra isometries could 
exist. This case is not empty as we can see by looking at the de Sitter 
space, exceptional in many respects. Clearly our approach  is based upon the 
existance of a preferred family of space sections, so it is bound to break 
down in the de Sitter case.               Fortunately the existance of
 invariant kernels in this particular spacetime  are well-known [\sit][\Sitt].

\bigskip
\noi      
The possibility of connecting  two different definitions of the 
one-particle space by a unitary transformation was  known previously 
[\parkful][\moren].  But here we have additionally proved that
(in the generic case)
 the connected group of spacetime isometries acts  unitarily in 
$ \hron ^+ = \bigoplus _n  \hron _n ^+$, whatever is the choice of admissible 
basis made in each   $\sron  _n $.

\noi                   This fact strongly suggests that we  
 treat all admissible choices  on equal footing.
In physical terms, it means that all possible definitions of the
 positive-frequency one-particle states could correspond to equivalent 
representations of the same physics.
In other words,  the tremendous arbitrariness involved in 
the formalism is  physically irrelevant, and can be seen
 as a sort of gauge freedom.

\medskip
\noi
 This  interpretation must be understood as an attempt to give a definition 
of particle {\it as much as possible\/}  independent of the observer.
It could be easily generalized to several other classes of spacetimes.

\noi
Of course the operational meaning of this picture remains to be investigated 
(for instance the presence of a horizon could perhaps  result in some
 limitations).
In our opinion, this task should be undertaken 
 only {\it after} Fock space construction is achieved.

\noi   When statically bounded expansion is considered, it becomes clear 
that our definition of a particle departs from the popular scheme of "in and 
out vacua". 
In sofar as detectors and observers should be invoked at this stage,
 our point of 
view amounts to think of a unique class of  observers,  rather
 than of two classes  of observers  respectively submitted to "in" and
 "out" asymptotic conditions.

\noi      In contradistinction, there is no conceptual discrepancy  between the 
interpretation proposed here and various efforts made in the past [\birl] 
and recently  [\pdv][\bel] in order to 
exhibit, under some circumstances,  a  unique, distinguished,
 "definition of the vacuum".   
 Indeed, in asmuch as  a distinguished definition 
can be actually  considered as satisfactory,
 it becomes a matter of taste to utilize this  representation 
 rather than any other one equivalent to it.

 \bigskip
\noi      The author is indebted to J.Renaud and P.Teyssandier for
 discussions about de Sitter spacetimes.

\bigskip
APPENDIX 1

\noi
It is convenient to define 
$ S = B^2$ 
and to utilize the cosmic time $\overline t$.
In this section, overlines will be provisionally dropped for typographical 
simplicity.

\noi    The calculation presented below is affected by the dimensionality of 
spacetime.
Although we are mainly interested by the case of $V_4$, 
applications in different contexts may be of interest.
Therefore, in this Appendix, let  
   $V_{q+1} =  {\bf R}   \times   V_q  $               be the spacetime 
manifold. Here  $i,j,k $ run from 1 to $q$, and Greek labels from $0$ to $q$. 
Parenthesis for indices means symmetrization.

\noi  The  argument is purely local;  compactness of $V_q$ plays no role.

\noi  Using the  cosmic  time coordinate, it is possible to cast the metric 
into the   following form
$$ ds^2 = dt^2 -  S(t)  \gam _{ij}   dx^i  dx^j      \eqno (A1)           $$
The point $\xi \in V_q $ has coordinates  $x^i$.
In these coordinates, $g _{00} =1,   \quad g_{0i} = 0 $ and  
$ g_{ij} = -S(t)  \gam _{ij} (\xi)  $.
Coordinates in $V-q$ are arbitrary untill we need to specify them.
Beware that here a dot means differentiation with respect to the comoving 
time. 

\noi
    For a Killing vector $X$ of $V_{q+1}$,  we are interested in developping
 the equation 
$$ \nab _\mu  X_\nu + \nab _\nu  X _\mu =0    \eqno (A2) $$
The question is whether $X^0$ may be different from zero.
 Christoffel symbols are as follows [\ave] 
 $$ {{ \Gamma _0}^0 } _0   =  {{ \Gamma _0}^0 } _i    =  
{{ \Gamma _0}^i } _0      =0           $$
$$  {{ \Gamma _i}^0 } _j   =  \half \partial _0  g_{ij}      \eqno (A3)    $$
$$ {{ \Gamma _0}^i } _j   =  \half g^{ik}  \partial _0  g_{kj}    
                                                              \eqno  (A4)   $$
But    $ g^{ij} = - S^{-1} \gam ^{ij}  $, hence 
$$ g^{ik}  \partial  _0    g _ {kj}  = 
            {  \dot {S}  \over  S  }  \delta ^i _ j                    $$
and finally
$$    {{ \Gamma _0}^i } _j = 
 {\dot {S} \over  2S}  \delta  ^ i  _j     \eqno  (A5)  $$
The remaining components of the affinity are
$$   {{ \Gamma _i}^ k }_ j  =    
{{ (\widetilde {\Gamma})   _i }^k } _j        \eqno  (A6)  $$
where $ \widetilde \Gamma $ is the affine connection for the time-depending
 tensor $ g_ {ij} $ defined on $V_3$.
Elementary manipulations show that    
$$ \widetilde \Gamma =  \overline \Gamma    \eqno  (A7) $$
 where $\overline \Gamma $ is the 
affinity of the three-metric   $ \gam $ of $V_3 $.

\noi   Let us now express the contents of (A2).
It is clear that $\nab _0  X _0 =0   $.
Hence 
       $$  \partial _0   X _ 0   = 0     $$
Thus $X_0 = X^0 $ depends only on $\xi$. Now, on the one hand we have
$$ \nab _0  X _i = \partial _0 X _i -    {{ \Gamma _0}^j } _i   X_ j   $$
From (A5) we find 
$$ \nab _0 X_i  =
 \partial _0 X_i  -  {\dot {S} \over  2S}  X _ i        \eqno (A8)  $$
On the other hand we have
 $$ \nab _i X_0 =    \partial _i    X_0
    -  {{ \Gamma _i}^j } _0   X_ j             $$
According to (A5)   we get 
                $$ \nab _i X_0 = \partial _i  X_0  
    - { \dot {S} \over 2S}  X_  i                 \eqno  (A9)        $$
The sum of (A8)(A9) and condition that $ \nab  _{(i}  X _ {0)} $ vanishes give 
$$   \partial _0  X_i   +  \partial _i   X_ 0   =
{\dot {S}  \over  S}  X  _ i                           \eqno (A10)  $$
Then, we must express that 
$ \nab _ {(i}         X_{j)}  $ vanishes.
$$ \nab _i  X _ j =  \partial  _i  X _ j     -  
 {{ \Gamma _i}^k } _j  X_k    -     {{ \Gamma _i}^0} _j   X_0        $$
By (A6)(A7)(A3)  
$$ \nab _i  X _ j =  \partial  _i  X _ j     - 
 {{ \overline {\Gamma} _i}^k } _j  X_k    + \half (\partial _0 g_{ij}) X_0  $$
where   $ \partial _0  g_ {ij}  =  - \dot {S}  \gamma _{ij}    $.
$$ \nab _i  X _ j =  \partial  _i  X _ j     -
 {{ \overline {\Gamma} _i}^k } _j  X_k    -
{\dot {S}  \over 2}  \gamma _ {ij}  X_ 0           \eqno (A11)  $$
So we must write
$$  \partial  _i  X _ j + \partial _j  X_i  -
2   {{ \overline {\Gamma} _i}^k } _j  X_k   -
\dot {S} \gam _{ij}   X_0     + 0                          \eqno (A12)     $$
Contracted multiplication of (A12) by $ \gam ^{ij} $ yields
$$  2  \gam ^{ij} \partial _i  X _ j  -
2 \gam ^{ij}   {{ \overline {\Gamma} _i}^k } _j  X_k   -
   q  \dot {S}  X_0   =  0    $$
(since space sections are q-dimensional).
So far the coordinates in $V_q$ have been left arbitrary.
Using now harmonic coordinates in  $V_q$, we cancel the second term in the 
above formula.
Hence    
$  2  \gam ^ {ij} \partial _i  X _ j   =  q  \dot {S} X_ 0        $
and after differentiation  with respect to $x^0$,
$$ 2 \gam ^{ij}  \partial _0  \partial _i X_j   =
            q  \ddot S  X_0                                   \eqno (A13)   $$
Now, differentiating (A10) with respect to the $j^{th}$ coordinate we obtain
$$ \partial _0 \partial _j   X _i  +      \partial _i \partial _j   X _0  =
{\dot {S}  \over S }  \partial _ j X _i              $$
Since $\partial _0 \gam ^{ij} = 0 $, we notice that 
$$ \partial _ 0 (\gam ^{ij} \partial _i X_j )   =
\gam ^{ij} \partial _ 0 \partial _i   X_j                             $$
hence the contraction
$$   \partial _0 ( \gam ^ {ij} \partial _i  X _j   )  +
\gam ^{ij} \partial _ j \partial _i   X_0     =
{\dot {S} \over S}   \gam ^{ij}   \partial _j   X _i          \eqno (A14)    $$
Insert (A13) into  (A14);  we finally get
$$ (\ddot {S} - { {\dot S}^2   \over S}  ) X_ 0
+  {2 \over q }    \gam ^{ij} \partial _ j \partial _i   X_0   =  0   $$
As the second term doesnot depend on time, it is necessary that 
$$       \ddot  {S } - {\dot {S}^2   \over S}    = const   \eqno (A15)  $$
It is noteworthy that this condition is independent of the dimension.
For a  flat and open  $V_q$, 
 the vanishing of the constant in the r.h.s. of (A15) 
 corresponds  to the steady-state universe  $S= const. \   e^{lt}$,
 with Hubble 
constant $l$, which covers a half of the de Sitter manifold.

\noi  For $q$-spherical  space sections 
(the case we are more specially concerned 
with in this paper) we have $\ddot S - \dot S^2 / S = 2 $,
satisfied by the de Sitter universe,             $ S =  \cosh ^2(t/l) $.

\bigskip
\noi
\centerline {\bf Notes and References}

               \medskip

\item {\Dew} B.S.De Witt and R.W.Brehme,  Ann.Phys. (NY) {\bf 9}, 220 (1960);
         B.S.De Witt,  {\it Dynamical Theory of Groups and Fields\/},
Gordon and Breach, New York (1965).

\item  {\licihes} A.Lichnerowicz, {\it Propagateurs et Commutateurs en 
Relativit\'e G\'en\'erale\/}, Inst. des Hautes Etudes Scientifiques,
 Publications  Math\'ematiques $N^o $ 10 (1961). 
About  isometries see   Sec. 6, p.16. 

 \item {\jap}        H.Nariai and K.Tanabe, 
Progress of Theoretical Physics, {\bf 55}, 1116-1132 (1976).
  R.R.Caldwell, 
Phys. Rev. D {\bf 48}, 4688-4692 (1993).

\item {\moren}   C. Moreno,  Reports on Math. Phys. {\bf 17}, 333 (1980)

\item  {\birl}  N.D.Birrell and P.C.W. Davies,
 {\it Quantum fields in curved space},
Cambridge University Press, (1982).

\item  {\wein} In the physical literature,
 they are often called "Weinberg two-point 
functions"  and introduced as expectation values of products of field operators 
in the vacuum.  In that context it was pointed out that in the simple case of 
an asymptotically static, spatially flat, Robertson-Walker model, the Weinberg 
function constructed using the "in vacuum" is invariant under spatial 
rotations and translations (Birrell and Davies, ref. [\birl], Chap.3,  p.58).
Our approach is radically different, for we first introduce the 
positive/negative-frequency kernel and plan to further define field operators 
with the help of it.

 \item  {\ouch}    A.Lichnerowicz,
 in "Relativity, Groups and Topology, Les Houches 1963"
De Witt and De Witt Eds. Science Publishers Inc. New York (1964).

\item  {\pdv} In the presence of curvature  the words "positive, negative 
frequency" do not refer to a periodic functional dependence, 
and must be understood in a loose  sense. See however:

\item  { \  } Ph.Droz-Vincent, Letters in Math. Phys. {\bf 36}, 277-290 (1996).
The main result obtained therein  requires  an assumption of periodicity and 
exclude solutions of unbounded growth in time.  In contradistinction, 
 calculations of its Sections 1,2  apply to all the FRW-like universes 
considered here.

\item  {\seg}  I.E.Segal, Journ. of Math. Phys. {\bf 1}, 468-488 (1960).

\item {\cheval} E.Combet, S\'eminaire Physique Math\'ematique, College de 
France (1965).
M.Chevalier, J.Math. Pures Appl. {\bf 53}, 223 (1974).                  
 C. Moreno,   Lett. in Math.Phys. {\bf 1}, 407 (1977);
Jour.Math.Phys. {\bf 18}, 2153 (1977).

\item  {\unit}  The author of ref.[\moren] was concerned by {\it real }
 solutions of the Klein-Gordon  
equation. Re-phrasing his result in the complex framework leads to replace 
"symplectic"  by  "unitary".

\item  {\sit} N.A.Chernikov and E.A.Tagirov, 
 Ann.  Inst.H.Poincar\'e, {\bf 9A}, 109 (1968).
 These authors  mainly consider {\it nonminimal coupling\/}.

\item  {\energ}     The kinetic energy $K$ must not be confused with the 
"timelike energy"   $p_0$  which is not conserved except in the trivial case 
of a static universe.

\item  { \   } In the particular example of a 
Friedman universe in the usual sense  (flat space sections),
the conservation of kinetic energy may be related with the constancy
 of the space components of momentum, which stems from space 
translation invariance.   But in general kinetic energy is conserved 
irrespective of a possible isometry group; it may happen that $V_3$ has no 
isometry at all, but the form (\frw) of the metric ensures that $K$ is 
conserved anyway.

   \item  {\Sitt}   In addition to ref. {\sit} see also 
Geheniau and  Ch. Schomblond (1968); Ch.Schomblond and  Ph.Spindel (1975);
B.Allen,  Phys.Rev.(1985).  These works are more specially concerned with 
the "steady-state" manifold.

\item {\berg} M.Berger, P.Gauduchon and E.Mazet,
 {\it Lecture Notes in Maths 194\/}, 
Springer-Verlag, Berlin, Heidelberg, New York  (1971);
S.Gallot, D. Hulin and J.Lafontaine, {\it Riemannian Geometry\/}, 
Springer-Verlag, Berlin, Heidelberg, New York  (1987) Chap. IV D, pp. 196-202.
                          
\item {\parkful}  When
 the space sections are 3-spheres, an alternative labelling 
explicitly  using spherical harmonics will be more familiar to the physicist: 
see for instance 
M.Bander and C.Itzykson, Rev. Mod. Phys. {\bf 38}, 346 (1966);
L.Parker and S.A. Fulling, Phys. Rev. D {\bf 9}, 341-354 (1974) and 
references therein.

\item {\bel}
  L. Bel in {\it "On Einstein's  Path"},
 a Festschrift to Engelbert Schucking on 
the occasion of his 70th birthday, Alex Harvey ed; Springer-Verlag (to appear).
 Up to now the results presented therein concern formal series solutions of 
the  wave equation.

\item  {\ave}  See for instance "Orthogonal Coordinates" in A.Avez,
Ann. Inst. H.Poincar\'e,  {\bf 1}, 291-300 (1964).   
                                                                       
\end